# Influenza Virus Vaccine Efficacy Based On Conserved Sequence Alignment

Baby Jerald A. #1
RIIC, Biocomputing Research Group,
Dayananda sagar Institutions
Bangalore, India
[1] babyjerald@gmail.com

T.R. Gopalakrishnan Nair*2
ARAMCO Endowed Chair – Technology, PMU, KSA.
RIIC, Dayananda Sagar Institutions
Bangalore, India
[2] trgnair@gmail.com

*Abstract*: The rapid outbreak of bird flu challenges the outcome of effective vaccine for the upcoming years. The recent research established different norms to eliminate flu pandemics. This can be made possible with skilled experimental analyses and by tracking the recent virulent strain and can be broadly applicable with effective testing of vaccine efficacy. Every year World Health Organization (WHO) reveals the administration of drug and vaccine to counter arrest the spread of flu among the population. As there are recurrent failures in priming the population, the complete eradication of the flu pandemic is still appears to be an unresolved problem. To overcome the current scenario, high level efforts with theoretical and practical research is required and it can enhance the scope in this field. The recent advancements also allow the researchers to endeavor effective vaccine to meet the emerging flu types. Only the standardized vaccination among the population at the time of flu pandemics will revolutionize the current propositions against influenza virus. This paper shows the deficiencies of vaccine fitness research as there are reported failures and less efficacy of vaccine even after priming the population from referred evidences and studies. It also shows simple experimental approach in detecting the effective vaccine among the vaccines announced by WHO.

## I. INTRODUCTION

The influenza A virus subtype H1N1 was widespread in the year 2009 and was observed first among Mexicans and Americans [1].The emergence of H1N1 pandemics presented a serious and highly complex public health challenge. During the rapid outbreak of pandemics, the potentially imminent availability of effective vaccine is found to be the best way of protecting people. However the efficacy of the vaccine and the protective effects of the vaccine are considered as the fundamental strategy on protecting the population from severe disease or death. The viral RNA mutates and develops resistance, which requires very careful consideration. Regardless, well monitored genomic study is strictly needed as the virus has the capacity to reassort between several species [2]. Through all these distinct features, virus is evolved in to human pathogen that is capable of causing a pandemic dimension. The Oseltamivir resistance of this particular virus [3] turned to be the more harmful, so that the virus would then be capable of easily transmissible among people resulting in illnesses related to H1N1 virus [4,5,6, and 7]. This work reminds the necessity of vaccine efficacy and also inadequate supply of vaccine in many scenarios.

The inequity in vaccine supply should be intensely monitored to ensure availability of the stockpile at the correct time. Such issues are threats to public and should be clarified at the earlier in order to protect the population from the flu pandemics for social welfare. In addition, one concerns about the effectiveness of the vaccine. In such situations extensive clinical trials are needed and should be supported by large safety databases to ensure the safety in priming the population. The ideology is that testing the efficacy of vaccine with sequence based method, which might be real or perceived and possibly reports vaccine efficacy.

## II. RESEARCH BACKGROUND

H1N1 virus found to be capable of causing sustained and widespread human to human infection [8] and the survey report says H1N1 is predominant in 5 of 24 years [9]. As with other influenza vaccines, the development of effective H1N1 vaccines is notable in 41 years as this strain found to cause severe deaths in both children and adults [10]. One of the studies shows the development of Transverse Myelitis in association with H1N1 vaccination which demonstrates the less efficacy of vaccine [11]. Likewise wide variability in the serological assays was observed while testing the efficacy of H1N1 vaccine. Similarly other vaccination study shows the outbreak of Guillian Barre Syndrome with respect to immunization [12]. As outlined in 2006 vaccination report in Israel campaign, it is clear that the death occurs 1 per 1000 in a week after immunization [13]. More over, it is been proved that H1N1 adjuvant vaccines led to more local symptoms [14, 15, 16, and 17]. In recent past, many studies show failures in safety and efficacy of H1N1 vaccine mainly among students and health care workers [18, 19, and 20]. As a result difficult decisions will probably need to be taken to address this threat. In fact, the vaccines should be designed with the purpose of stimulating immune response and should possess small antigenic distance related to the concurrent virulent strains to elicit the immune response [8].

However, the vaccine also can be designed based on sequence similarity between the vaccine strain and the concurrent viral strain. As Subunit vaccines for the seasonal influenza is very safe and well tolerated, the matching of neuraminidase protein sequence with that of vaccine strain is been examined for testing the efficacy of vaccine.

### III. METHODOLOGY

The viral strain used in this study is (A/India/NIV1028798/2010(H1N1) and the sequence information is obtained from NCBI - Influenza Virus Resource. (www.ncbi.nlm.nih.gov/genomes/FLU/FLU.html). Sequence alignment technique is used to find the protein match between the given sequences, perhaps extending them to meet specific constraints. It is an essential tool for many biological researches and is implemented here to evaluate the vaccine efficacy.

Table 1 gives the information about the Viral strain, Vaccine Virus and the length of the sequence. It also gives the exact matching score between the Vaccine and Viral sequence.

The entire vaccine sequences recommended for vaccination were retrieved from the WHO website (http://www.who.int/influenza/vaccines/virus/recommendations/2012south/en/) and the fasta sequences of the respective vaccine virus was downloaded from NCBI for further analysis. The retrieved fasta sequences of all the vaccine virus was subjected to alignment with the Influenza A virus sequence (A/India/NIV1028798/2010(H1N1). The Genbank ID of the Indian virulent virus selected for this evaluation is AEM63448.1 which was collected from the Influenza virus resource (NCBI). The publicly available tool of European Molecular Biological Laboratory (EMBL-EBI) is utilized to align the sequences using CLUSTAL W.

Table 1: Pairwise sequence Alignment Scores between the Indian Human Viral strain protein Neuraminidase and the Vaccine viruses.

| S.No | Sequence A | Sequence B (Vaccine viruses) | Length A | Length B | Score (in %) |
|---|---|---|---|---|---|
| 1. | Influenza A virus (A/India/NIV1028800/2010 H1N1) | >gi|227809834|gb|ACP41107.1| neuraminidase [Influenza A virus (A/California/04/2009(H1N1))] | 566 | 489 | 98.0 |
| 2. | Influenza A virus (A/India/NIV1028800/2010 H1N1) | >gi|321272925|gb|ADW80519.1| neuraminidase [Influenza A virus (A/Perth/16/2009(H3N2))] | 566 | 469 | 42.0 |
| 3. | Influenza A virus (A/India/NIV1028800/2010 H1N1) | >gi|307101756|gb|ADN32819.1| neuraminidase [Influenza B virus (B/Brisbane/60/2008)] | 566 | 466 | 28.0 |

### IV. RESULTS AND DISCUSSION

As compared with other influenza virus, H1N1 shows antigenic heterogenicity and can further evolve in to the most virulent strain with the progress of time. As a result, development of vaccine that is capable of inducing cross protective immunity against the current circulating and future emerging strains is vitally important. For this reason, the use of sequence alignment technique helps to identify the best matching protein sequence. The preposition of best matching score helps to predict the high affinity vaccine. The derived vaccine strain with high potentiality can be used with effective clinical trials. It is anticipated that if H1N1 evolves in to a pandemic strain in future, the current H1N1 vaccine strain will not be a perfect match. In this situation the application of rapid and simple techniques is needed to evaluate an effective vaccine. The strength of this study is to show the simple protocol in testing the analytical effectiveness of vaccine which could be a preliminary estimation.

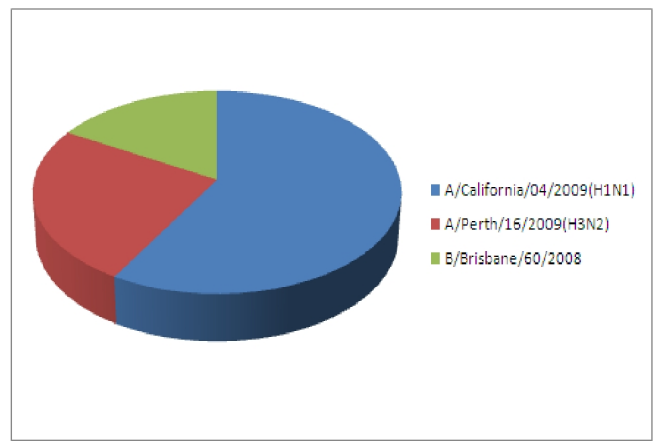

Fig 2: Pie Chart Representation of effective vaccine based on Sequence Similarity Alignment between vaccine and viral strain.

There is no certainty about the rate of success when the H1N1 viruses continue to drift. To overcome such difficulties, the tracking of previous records of pandemics should be maintained and also simple protocols like sequence alignment can be done to test the efficacy of vaccine against future H1N1 virulent strains. This might results in increased cross reactivity as the antigenic distance decreases and can able to provide protection against future H1N1 viruses. The efficacy of vaccine among the human is quantitatively presented (Fig 2) to show the percentage of similarity alignment.

### V. CONCLUSION

In this paper, we present a non-rigid approach to jointly solve the tasks of effective vaccine determination using computational methods for which one may not find complicate in applying this strategy at the time of outbreaks. The experimental result highlights the possibility of vaccine efficacy determination and preferential usage out of all the vaccines announced by WHO this year on several pose estimation and probabilistic scenarios.

As presented above, the evaluation of effectiveness of vaccine through this approach will reduce the failure rate of vaccines. This test is applicable to all variants of Influenza Virus and it involves rapid and simple procedures as there is no need of expensive equipments. It is appropriate to perform this test before the outbreak of pandemics so as to bring down the spread of flu at the time of pandemics. Although three vaccine strains have been tested in this study, it is shown that the strain A/California/01/2009 (H1N1) is more effective when compared with the other two strains. Such potential vaccination among the population inherently increases the immune responses than priming the population with A/Perth/16/2009 (H3N2) and B/Brisbane/60/2008. The vaccine derived with high efficacy is very safe for seasonal influenza. Sequence alignment of capsid proteins either hemagglutinin or neuraminidase with that of vaccine strains increase the level of uncertainty surrounding the risk of rare unanticipated adverse effects.